\documentclass{emulateapj} \usepackage{apjfonts,psfig} 
 
\def\cm{{\rm\,cm}}

\def\kms{{\rm\,km\,s^{-1}}}

\def\kpc{{\rm\,kpc}}

\def\ergpscm{{\rm\,erg\,s}^{-1}{\cm}^{-2}}

\def\ntimes{\!\times\!}
\def\spose#1{\hbox to 0pt{#1\hss}}
\def\lta{\mathrel{\spose{\lower 3pt\hbox{$\mathchar"218$}}
     \raise 2.0pt\hbox{$\mathchar"13C$}}}
\def\gta{\mathrel{\spose{\lower 3pt\hbox{$\mathchar"218$}}
     \raise 2.0pt\hbox{$\mathchar"13E$}}}
\def\etal{{\sl et~al.\ }}

\slugcomment{Accepted for ApJL, 27-01-2005} 
 
\shorttitle{Detection of Planetary Nebulae in the Coma Cluster}
\shortauthors{Gerhard, O. et al.} 
 
\begin{document} 
 
 
\title{Detection of Intracluster Planetary Nebulae in the Coma
  Cluster\footnote{Based on data collected with the FOCAS spectrograph
    at the Subaru Telescope, which is operated by the National
    Astronomical Observatory of Japan, during observing run S04A-024.}}
 
 
\author{Ortwin Gerhard\altaffilmark{1},  
        Magda Arnaboldi\altaffilmark{2},  
        Kenneth C. Freeman\altaffilmark{3}, \\
        Nobunari Kashikawa\altaffilmark{4}, 
        Sadanori Okamura\altaffilmark{5}, 
        Naoki Yasuda\altaffilmark{6}} 
\altaffiltext{1}{Astronomisches Institut der Universitaet, CH-4102 
  Binningen, Switzerland (ortwin.gerhard@unibas.ch)} 
\altaffiltext{2}{INAF, Oss. Astr. di Torino, Strada Osservatorio 20, 10025 
        Pino Torinese, Italy (arnaboldi@to.astro.it)}
\altaffiltext{3}{RSAA, Mt. Stromlo Observatory, Cotter Road, Weston Creek, ACT 
        2611, Australia (kcf@mso.anu.edu.au)}
\altaffiltext{4}{NAOJ, 2-21-1 Osawa, Mitaka, Tokyo, 181-8588, Japan
        (kashik@zone.mtk.nao.ac.jp)}
\altaffiltext{5}{Dept. of Astronomy and RESCEU, School of Science, The
       Univ. of Tokyo, Tokyo 113-0033, Japan (okamura@astron.s.u-tokyo.ac.jp)}
\altaffiltext{6}{Institute for Cosmic Ray Research, Univ. of Tokyo,
        Kashiwa, Chiba 277-8582, Japan (yasuda@icrr.u-tokyo.ac.jp)}
  
\begin{abstract} 
  [OIII]$\lambda5007$\AA\ emission lines of 16 intracluster planetary
  nebulae candidates in the Coma cluster were detected with a
  Multi-Slit Imaging Spectroscopy (MSIS) technique using FOCAS on the
  Subaru telescope. The identification of these faint emission sources
  as PNe is supported by (i) their point-like flux distribution in
  both space and wavelength, with tight limits on the continuum flux;
  (ii) the identification of the second [OIII]$\lambda4959$ line in
  the only object at high enough velocity that this line too falls
  into the filter bandpass; (iii) emission line fluxes consistent with
  PNe at 100 Mpc distance, in the range $2.8 \ntimes 10^{-19} -
  1.7\ntimes 10^{-18} \ergpscm$; and (iv) a narrow velocity
  distribution approximately centered on the systemic velocity of the
  Coma cluster. Comparing with the velocities of galaxies in our
  field, we conclude that the great majority of these candidates would
  be intracluster PNe, free floating in the Coma cluster core. Their
  velocity dispersion is $\sim 760\kms$, and their mean velocity is
  lower than that of the galaxies. The velocity distribution suggests
  that the intracluster stellar population has different dynamics from
  the galaxies in the Coma cluster core.
\end{abstract} 
 
 
\section{Introduction} 
Cosmological simulations of structure formation predict that galaxies
are dramatically modified by galaxy interactions during the assembly
of galaxy clusters, losing a substantial fraction of their stellar
mass which today must be in the form of intracluster stars (Murante
\etal\ 2004).  Observations now show that there is a substantial
amount of {\it intracluster stellar population}, which is observed as
diffuse intracluster light (ICL, Bernstein et al.\ 1995), or as
individual stars, {\it i.e.,} planetary nebulae (PNe; Arnaboldi et
al.\ 2003, Feldmeier et al.\ 2004) and red giant stars (Durrell et
al.\ 2002). The ICL represents about 10\% of the stellar mass in the
Virgo cluster and as much as 50\% in rich Abell clusters (Arnaboldi
2004).

Intracluster planetary nebulae (ICPNe) are the best suited tracers for
dynamical studies of the ICL because of their strong [OIII] 5007\AA\
emission, which allows both identification and radial velocity
measurement. Measuring the projected phase-space for the ICPNe
constrains  how and when this
light originated (Napolitano et al.\ 2003). So far, ICPNe velocities
have been measured only in the Virgo cluster at 15 Mpc distance
(Arnaboldi \etal\ 2004).
Here we extend these studies to the Coma cluster (Abell 1656),
one of the best studied nearby galaxy clusters. Of these,
it is the richest and the most compact one, providing a
laboratory of prime importance for studying the effect of a dense
environment on galaxy evolution. The ICL in Coma
has been suspected since the early study by Zwicky in 1951 and was
confirmed by Thuan \& Kormendy (1977) and Bernstein \etal\ (1995).
 The brightest PNe in the Coma cluster at distance of $\simeq 95 {\rm
 Mpc}$ (Bernstein \etal\ 1995) have fluxes of $2.2\times 10^{-18}
\ergpscm$ (Ciardullo \etal\ 2002). To detect even the
brightest of the Coma PNe, we must find a way to decrease
substantially the noise from the night sky.

We can achieve this with a technique that is similar to the approach
used by Tran et al.\ (2004) and Martin \& Sawicki (2004) to search for
Ly$\alpha$ emitting galaxies at very high redshift. It combines a mask
of parallel multiple slits with a narrow band filter, centered around
the [OIII]$\lambda5007$\AA\ line at the redshift of the Coma cluster,
to obtain spectra of all PNe that happen to lie behind the
slits. Because the [OIII] emission lines from PNe are only a few km
s$^{-1}$ wide, their entire flux still falls into a small number of
pixels in the 2D-spectrum, determined by the slitwidth and seeing. On
the contrary, the sky emission is dispersed in wavelength, allowing a
large increase in signal-to-noise (S/N). The narrow band filter limits
the length of the spectra on the CCD, so that more slits can be
exposed. For brevity we will refer to this as the Multi
Slit Imaging Spectroscopy (MSIS) technique.  No conventional imaging
technique can decrease the sky surface brightness in a
similar way.  In this Letter we show that with this technique we are
indeed able to detect PNe in the Coma cluster, and briefly describe
the first results obtained from one MSIS image in Coma.

\section{Observations} 

The observations were carried out with the FOCAS spectrograph at the
8.2m Subaru telescope on April 23, 2004.  The spatial resolution of
FOCAS is $0''.1$pix$^{-1}$, so the $6'$ diameter of the circular field
of view (FOV) corresponds to 3600 pixels.  We used grating 300B, which
gives a measured dispersion of 1.4--1.5\AA\ per pixel on the two FOCAS
CCD chips.  We used the N512 filter with FWHM of 60\AA, centered at
$\lambda_c=5121$\AA, the wavelength of the redshifted [OIII] emission
from a PN at the mean velocity of the Coma cluster.  The FWHM includes
only $\pm 1.6\times$ the galaxy velocity dispersion in the Coma
core. We will be able to detect the redshifted [OIII] 4959\AA\
emission only for the brighter PNe with the largest receding
velocities, $\gta 7400\kms$ , so that it can be seen near the blue
edge of the filter.

The light passing through the narrow band filter and a $0''.6$-wide
long slit, and dispersed by the grism projects down to a spectrum of
about 43 pixels. A mask was therefore constructed with uniform long
slits spaced every 50 pixels, and interrupted only by short sections
to ensure mechanical stability (see Fig.~1).  The slitlet width on the
mask was 0.$''$6, corresponding to six 0.$''$1 pixels. The area
surveyed by this mask configuration is then about $(3600/50)$ slits $
\times 0''.6 \times 360'' \times (\pi/4) = 12215$ arcsec$^2$, or
12\% of the whole FOCAS FOV.

With this mask, grism, and filter we took six 30 min exposures of a
field centered at $\alpha=$12:57:17, $\delta=$+28:09:35 (B1950). This
field is near the X-ray centroid of the Coma cluster and is
essentially concident with the field observed by Bernstein et al.\
(1995).  The measured seeing of the images varied from 0.$''6$ to
0.$''$8.  The CCD readout of the image was done with full spectral
resolution and with a spatial binning of 2 pixels.  Monochromatic,
point-like emitters appear as elongated ellipses on the CCD, with
width of approximately 4 rebinned pixels and height of 5 pixels in the
wavelength direction. Then the effective spectral resolution is
$\simeq 7.3$\AA, or $440\kms$. Data reduction was carried out in IRAF,
with flux calibration using the spectrophotometric standard star
BD+33d2642.

\begin{figure} 
\epsscale{0.9} 
\plotone{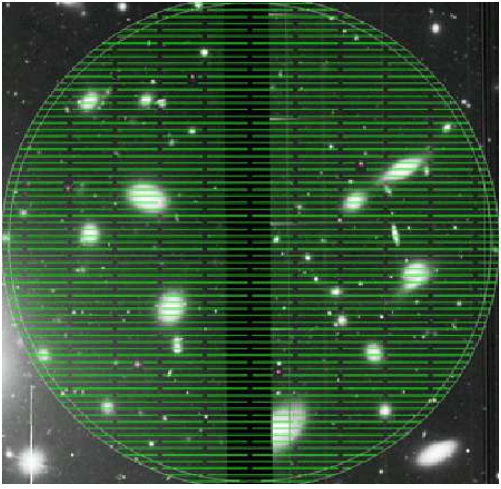} 
\caption{Multi-slit mask used in the observations, superposed on
the observed field. North is to the left and East is up. The extended
light at the NNW margin is from the halo of NGC 4874. The diameter of
the FOCAS FOV, indicated by the circles, is $6'$ (about 165 kpc at
Coma). The dispersion direction is West-East. Slitlets are arranged
along a total of 70 long slits, but with short interruptions for
mechanical stability of the mask.
\label{fig1}}
\end{figure}

\subsection{How many PNe do we expect per slit configuration?} 

We have estimated the average surface brightness of the diffuse light
in our Coma core field from the results of Bernstein et al.\ (1995) to
be $R= 24.7$ mag arcsec$^{-2}$. Assuming B-R = 1.0, the average
surface brightness in B in this field is 25.7.  The total magnitude in
the FOCAS circular field of $6'$ diameter is then $B_{TOT} = 13.18$,
which amounts to a total luminosity of $ L_B = 7.6 \times 10^{10}
L_\odot$. To estimate the corresponding number of PNe, we
assume that the planetary nebulae luminosity function (PNLF) can be
described by the formula of Ciardullo \etal\ (2002) which is a good
fit to numerous observations in nearby galaxies.  With the
observations described below, we will be complete for PNe $\sim 1.4$
mag fainter than the bright cutoff of the PNLF.  Using the
luminosity-specific PN density determined for an evolved population
such as in M87, $\alpha_{1.0,B} = 5.6\times 10^{-9}$ PN
L$_{B,\odot}^{-1}$ (Jacoby \etal\ 1990, Peletier
\etal\ 1990, Ciardullo \etal\ 1998), we expect $\sim 425$ PNe
associated with the diffuse light in this field, down to 1 mag fainter
than the cutoff. 

The fraction of the FOCAS FOV surveyed by the mask used in our
observations is 12\% (see Section 3), giving about 50 PNe located
behind the mask slitlets. However, because the seeing FWHM is nearly
equal to the slit width, of order half of these will have a
significantly lower flux measured than their true flux; for
FWHM=slitwidth, a point source at the center (edge) of the slit is
dimmed by 24\% (51\%). Also, we lose a small fraction of the remaining
PNe due to the limited filter bandwidth, if these PNe have the same
velocity distribution as the Coma galaxies.  Thus finally we expect to
detect approximately 20--30 PNe per mask.

\subsection{The influence of observational parameters 
on the detectability of PNe}\label{S/N}

Here we show how the S/N of the detected PNe depends on the
observational parameters, mainly seeing, slit-width, and spectral
resolution. For a distance modulus of 34.9, a PN at the bright end 
of the PNLF has a flux in the [OIII] 5007\AA\ line of
$
F_{5007,{\rm cutoff}} = 2.2 \times 10^{-18}\,\mbox{erg sec$^{-1}$ cm$^{-2}$}.
$ 
$0.''6$ resolution corresponds to 300 pc at this distance.

For a sky surface brightness $\mu_V {\rm mag}/{\rm arcsec}^{2}$, the
magnitude of sky in a pixel $\sigma_x''\ntimes \sigma_y''$
$ m_{v,sky} = \mu_V - 2.5 \log(\sigma_x\sigma_y)$, and the 
monochromatic flux from sky at 5500\AA\ in the pixel is
$ \log F_{sky,\lambda} = -0.4 m_{v,sky} - 8.4$, or $ F_{sky,\lambda} =
4 \ntimes 10^{-9} 10^{-0.4\mu_V}
\sigma_x\sigma_y\, \mbox{erg sec$^{-1}$ cm$^{-2}$\AA$^{-1}$}.
$
For a dispersion $d$ and slit width $\theta''$, the true spectral 
resolution is
$ d_{true} = d \kappa (\theta/\sigma_y)$, where $\kappa$ is a 
constant of order unity. 
In our observational set-up, the CCD is rebinned $2 \ntimes1$, giving a spatial
resolution $\sigma_x''\ntimes\sigma_y'' = 0''.2\ntimes0.''1$,
$d=1.45$\AA$/{\rm pixel}$,
$\theta''=0.''6$, and $\kappa=5/6$.
Thus the flux from the sky surface 
brightness in the rebinned FOCAS pixel is
$
f_{sky,pix} = 4 \ntimes 10^{-9} 10^{-0.4\mu_V} 
\sigma_x\sigma_y d_{true}\,\mbox{erg sec$^{-1}$ cm$^{-2}$}.
$
For our instrumental set-up, we have measured on the flux-calibrated
two-dimensional spectra
$
f_{sky,pix} = 7.7 \ntimes 10^{-19}\,\mbox{erg sec$^{-1}$ cm$^{-2}$},
$ 
which gives an effective $\mu_V=22.2$ at 5120\AA, inserting our pixel 
sizes and spectral resolution.

The PN as a point-like, monochromatic source falls onto 
$n\simeq\kappa'\pi\phi^2/(4\sigma_x\sigma_y)$ pixels, where in
our observations the seeing is $\phi = $ FWHM $=0''.6$ and $\kappa'=1.4$.
We can thus compute the total signal-to-noise for a PN source.
For a total integration time
of 3 hrs, i..e. $t_{exp} = 1.08\ntimes 10^4$ sec, a total telescope
area of $ S_{tel} = 5.281\ntimes 10^5$ cm$^2$, and an overall
efficiency of telescope + spectrograph + airmass, $\epsilon = 0.1$,
and $h\nu = 3.61\ntimes 10^{-12}\, \mbox{erg = E}_{5500}$ at $\lambda =
5500$\AA\ we obtain
\newcommand{\SNRPN}{ {\rm SNR}_{PN} }
\begin{equation}
 \SNRPN = 7.0 \times 10^{-0.4\Delta m}
\end{equation}
$$ \times
\left(\frac{d}{1.45{\mbox \AA}/{\rm pix}}\right)^{-1/2}
\left(\frac{\phi}{0.''6}\right)^{-1}
\left(\frac{\theta}{6\sigma_y}\right)^{-1/2}
\left(\frac{t_{exp}}{3~hrs}\right)^{1/2}
$$
for a PN fainter than the cutoff by $\Delta m$
magnitudes.
Relative to the sky noise in one pixel, the S/N of a PN at the PNLF
cutoff $(\Delta m=0)$ would be 32. Both values assume that the
emission from the PN falls through the slit completely.
Detection of a source is normally considered to be secure if the S/N
relative to the sky noise in one pixel is $\ge 9$. In our data this
corresponds to $\SNRPN \ge 2$, i.e., $95\%$ probability of the source
being real. If we use this criterion for secure detection, we can
detect PNe about 1.4 magnitudes down the PNLF.

Equation (1) shows that independent of all other parameters, good
seeing is critical for these observations. Also the $\SNRPN$ increases
like the inverse square root of the true spectral resolution. However,
so that the light of a PN falls through the slitlet completely, we
must have $\theta\ge\phi$.  Because large slitwidths degrade the
spectral resolution, there exists a trade-off between $\SNRPN$ and the
number of PNe that can be observed throught the mask (previous
subsection).  Consider that the PNLF has a sharp cutoff and then it is
nearly flat. Therefore, as long as the $\SNRPN$ remains large enough
that we can detect PNe one magnitude fainter than the bright cutoff,
it is advantageous to use larger slit-widths. For with a larger
slit-width, the additional PNe that come through the mask outnumber
the PNe that are lost at the faintest fluxes. Typically, to observe
one slit configuration for $5-6.5$ hours will work as long as the
seeing $\phi$ is not worse than $0.''8$.

\begin{figure} 
\includegraphics[angle=90,scale=.45]{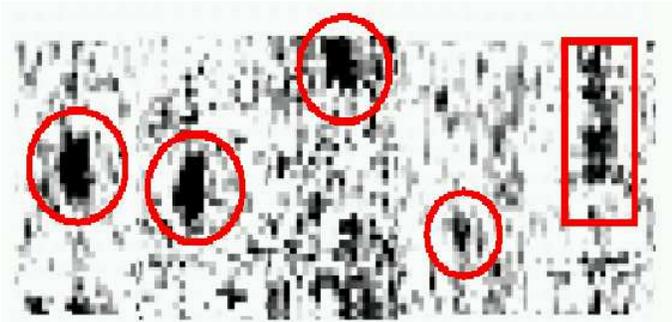} 
\caption{Two-dimensional median-averaged spectra of
emission objects in the FOCAS field. Wavelength is along the vertical
axis (507.7--514.15 nm) with a rebinned resolution of 1.5\AA\ per
pixel; the true spectral resolution is 7.3\AA\ or $440\kms$. The
horizontal direction is along the mask slitlets, with a rebinned
resolution of $0.''2$ per pixel.  The left four panels show PNe
candidates from the brightest to one of the faintest in the field.
The fluxes are ($34.7, 31.9, 18.6, 5.6$) ADU, corresponding to $(17,
16, 9, 3) \ntimes10^{-19}\ergpscm$. The rightmost panel shows the
spectrum of a background galaxy with continuum and strong absorption,
probably bluewards of Ly$\alpha$, and a possible line emission,
probably Ly$\alpha$.
\label{fig2}}
\end{figure}

\begin{figure} 
\epsscale{1.} 
\plotone{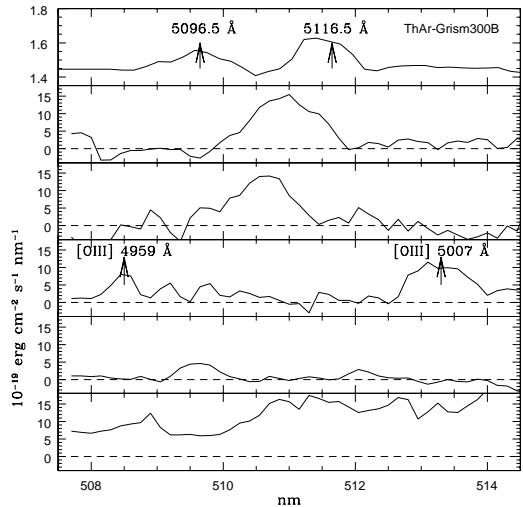} 
\caption{One-dimensional spectra of the PNe candidates and of the 
probable Ly$\alpha$ galaxy shown in Fig.~2 (same order from bottom to
top as in Fig.~2). The top panel shows the arclamp image with the 
two emission lines visible in our spectral range, observed with the
same set-up. The widths of the arclamp lines are very similar to the
widths of the brighter emission lines of the PNe candidates, showing
that these are unresolved in wavelength. The object in the third panel
from below is sufficiently redshifted that both lines of the oxygen
doublet [OIII]$\lambda4959$\AA, $\lambda5007$\AA\ fall in the filter
bandpass. The S/N of the emissions from the four PN candidates 
shown are ($20, 18, 11, 3.3 $) with respect to the noise per pixel
of 1.7 ADU or $8.5\ntimes 10^{-20}\ergpscm$.
\label{fig3}} 
\end{figure}

  \section{Results and Discussion: PNe in the Coma Cluster}

A total of 16 spectrally unresolved, point-like sources with
undetected continuum were detected on the two CCD chips. Their
emission line fluxes were in the range $2.8 \ntimes 10^{-19} - 1.7
\ntimes 10^{-18} \ergpscm$.  Objects with a flux of less than $2.5 \ntimes
10^{-19}\ergpscm$ (5 ADU) were considered as not real. Figure~2 shows
two-dimensional spectra of the two brightest objects, an intermediate
flux source, and a low S/N emission. These are all PN candidates.
Fig.~2 also shows a source with a possible emission line and a
continuum redwards of the line, but no continuum on the blue side,
most likely a Ly$\alpha$-absorbed background galaxy.

We now consider the evidence that the unresolved emission sources are
in fact planetary nebulae. Figure~3 shows their 1-d spectra,
obtained by summing the respective columns in the
2-d-spectra. The intermediate flux source in Fig.~2 is at sufficiently
large recession velocity that, if the observed line is
[OIII]$\lambda5007$\AA, the second $\lambda4959$\AA\ line also
falls into the narrow band filter. In fact, its 1-d-spectrum in Fig.~3
does show both lines of the doublet: the ratio of the equivalent
widths of both lines is consistent with the theoretical value of 1:3.

The brightest object has measured counts of 34.7 ADU in the median
image, corresponding to a total line flux of $1.7 \ntimes
10^{-18}\ergpscm$, and is detected at S/N$=20$. The faintest PN
candidate has a S/N$=3.3$, a line flux of 5.6 ADU, corresponding to
$2.8 \ntimes 10^{-19}\ergpscm$, or a total of 68
[OIII]$\lambda5007$\AA\ photons in three hours, about one photon every
three minutes.  The flux from a PN in Coma at the bright cutoff of the
PNLF is $2.2\times 10^{-18} \ergpscm$ (Section 2)\footnote{In the
[OIII] PNe magnitudes defined by Jacoby (1989) these two objects have
m$_{5007}=30.7$ and m$_{5007}=32.5$, and the bright cut-off of the
PNLF in Coma is at $m_{5007,{\rm cutoff}} = -4.5 + 34.9 =
30.4$.}. Note that our emission sources are generally not centered in
the slitlets, and even if they are, we lose 24\% (38\%) of the flux,
given the slitlet width of 0.$''$6 and a seeing FWHM of $0.''6$
($0.''8$).  Thus the brightest fluxes we measure are consistent with
PNe in Coma at the cutoff of the PNLF.

Our candidates are also undetected in the continuum.  The
$1\sigma$-upper limit on the pixel flux for the extracted spectra in
the range $5070$\AA$-5140$\AA, excluding the line, is 3.4 ADU,
corresponding to a continuum flux limit of $1.6\ntimes
10^{-20}\ergpscm\mbox{\AA}^{-1}$. Our brightest object thus has
EW$>$110\AA. The V-band continuum flux of an O8 star at the distance of
Coma is $4.3\ntimes 10^{-21}\ergpscm\mbox{\AA}^{-1}$ -- with our
observing technique we would see the continuum from a few O stars in
the Coma cluster!  This rules out compact HII regions such as or
brighter than those observed in Virgo (Gerhard \etal\ 2002).
Similarly, background galaxies would often have a much larger
continuum flux than our limit; for example, see the background object
shown in Fig.~2. For our brightest candidates, the large equivalent
widths are larger than those of known [OII] emitters at $z=0.37$ (Hogg
\etal\ 1998).  The line fluxes we measure are 10 - 100 times fainter,
and the surface number density is 10 - 100 times higher, than for
currently studied Ly$\alpha$ emitters at $z\sim 3$ (Fynbo \etal\
2003). Moreover, in the flux decade brighter than our brightest line
flux, we see no object without measurable continuum, and only one
object with blue continuum, for which EW=50\AA.

\begin{figure}
\epsscale{0.9}
\plotone{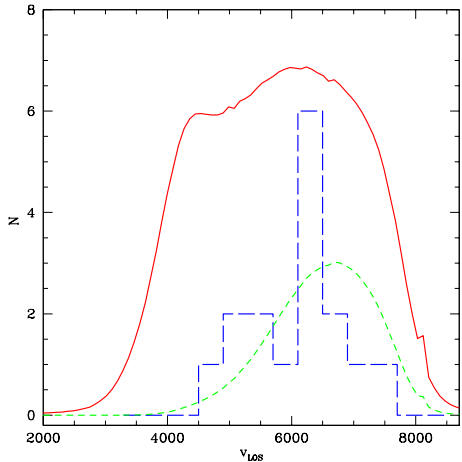} 
\caption{Velocity histogram for all 16 PNe candidates in the mask.
Mean and median velocities are $\overline{v}=6100\kms$ and $v_{\rm
med}=6224\kms$, respectively, and the velocity dispersion is
$\sigma=760\kms$ and $\sigma_{\rm med}=562\kms$ with respect to the
median. Overplotted is the measured filter transmission curve in our
observing set-up (full red line), using the continuum light from the
spectrophotometric standard star BD+33d2642, and (dashed curve) this
function times a Gaussian with $\overline{v}_{\rm Coma}=6853 \kms$ and
$\sigma_{\rm Coma} = 1082 \kms$ as for the galaxies (Colless \& Dunn
1996).
\label{fig4}} 
\end{figure}

Figure 4 shows the velocity histogram for our PNe candidates. They
have velocities between $4800-7560\kms$. With our instrumental setup,
we detect wavelengths in the range $5071-5135$\AA\ (FWHM), or
velocities in $3830-7680\kms$. The velocity histogram is clearly not
uniform in the filter bandpass (Fig.~4); an unclustered population of
background emission line galaxies would appear as such a nearly
uniform distribution, because only the faintest objects near the
filter edges would fall out of the sample.

In summary, we have the following evidence that our unresolved
emission sources are PNe in the Coma cluster: (i) They are unresolved
both spatially and in wavelength. (ii) They have no detectable
continuum, to a limit of $1.6\ntimes 10^{-20}\ergpscm\mbox{\AA}^{-1}$.
The brightest object thus has EW$>110$\AA. (iii) We have seen both
lines of the [OIII] doublet in the only source at sufficiently large
recession velocity that also $\lambda4959$\AA\ is redshifted into the
wavelength range probed. (iv) The emission fluxes of the brightest
objects are consistent with those of the brightest PNe in a population
at distance 100 Mpc. (v) The number density of our candidates is
consistent with that expected from the measured surface brightness of
the ICL in our Coma field. (vi) The distribution of recession
velocities is centered around the Coma cluster and is inconsistent
with a population of background objects uniformly distributed in
velocity. However, we cannot rule out that a population clustered in
velocity of so far undetected low-luminosity background emission
galaxies could contribute to our sample, if these exist in sufficient
numbers at $z\sim 3$ (Ly$\alpha$) or $z\sim 0.37$ ([OII]).

From Bernstein \etal\ (1995), approximately half of the total
luminosity in our field is in the ICL component of the Coma cluster,
the other half is in the brighter galaxies, while faint galaxies do
not contribute significantly.  In the PNe, we expect a larger diffuse
light fraction: PNe near bright galaxies will be harder to see because
of the increased background. We have superposed our PNe candidates on
the image of the field and compared their velocities with those of the
brighter galaxies in the field. None of our candidates is close in
space and velocity to any of these, but two PNe have measured
velocities and positions that could be consistent with the cD envelope
of NGC 4874, which is just outside our field, $\sim 120\kpc$ from the
field center.  NGC 4874 itself has a radial velocity of $7224 \kms$
and a velocity dispersion of $280\kms$ (Smith \etal\ 2000). Most of
our PNe, which have $\overline{v}=6100\kms$ and $\sigma=760\kms$, are
therefore not associated with the halo of NGC 4874.

Thus, the great majority of the detected PNe candidates would be
intracluster PNe (ICPNe) in the Coma cluster.  It is interesting that
the measured mean and dispersion velocities for these ICPNe are
smaller than those measured for the Coma cluster core centered around
NGC 4874, $\overline{v}_{\rm Coma}=6853 \kms$, velocity dispersion
$\sigma_{\rm Coma} = 1082 \kms$ (Colless \& Dunn, 1996). A KS test
shows that the ICPN velocity distribution has probability 12\% to be
drawn from the expected velocity distribution of galaxies in our
filter (Gaussian times filter band transmission).  This suggests that
the dynamics of the intracluster population in the Coma core may
differ from that of the galaxies, and calls for further investigation.


\acknowledgments 
We acknowledge financial support by the Swiss SNF and by INAF. We 
are grateful to the on-site Subaru staff for their support.

 

\end{document}